# Spontaneous Electric Polarization in Graphene Polytypes

Simon Salleh Atri[1], Wei Cao[2], Bar Alon[1], Nirmal Roy[1], Maayan Vizner Stern[1], Vladimir Falko[3], Moshe Goldstein[1], Leeor Kronik[4], Michael Urbakh[2], Oded Hod[2], Moshe Ben Shalom[1]

[1] School of Physics and Astronomy, Tel Aviv University, Tel Aviv, Israel

[2] Department of Physical Chemistry, School of Chemistry, The Raymond and Beverly Sackler Faculty of Exact Sciences and The Sackler Center for Computational Molecular and Materials Science, Tel Aviv University, Tel Aviv, Israel.

[3] National Graphene Institute, University of Manchester, Booth Street East, Manchester M13 9PL, UK

[4] Department of Molecular Chemistry and Materials Science, Weizmann Institute of Science, Rehovoth, Israel

**A crystalline solid is a periodic sequence of identical cells, each containing one or more atoms. If the constituting unit cell is not centrosymmetric, charge may distribute unevenly between the atoms, resulting in internal electric polarization[1]. This effect serves as the basis for numerous ferroelectric, piezoelectric, and pyroelectric phenomena[2]. In nearly all polar materials, including multilayered van der Waals stacks that were recently found to exhibit interfacial polarization, inversion symmetry is broken by having two or more atomic species within the unit cell[3–11]. Here, we show that even elemental crystals, consisting of one type of atom, and composed of *non-polar* centrosymmetric layers, exhibit electric polarization if arranged in an appropriate three-dimensional architecture. This concept is demonstrated here for inversion and mirror asymmetric mixed-stacking tetra-layer polytypes[12] of non-polar graphene sheets. Furthermore, we find that the room temperature out-of-plane electric polarization increases with external electrostatic doping, rather than decreases owing to screening. Using first-principles calculations, as well as tight-binding modeling, we unveil the origin of polytype-induced polarization and its dependence on doping. Extension of this idea to graphene multilayers suggests that solely by lateral shifts of constituent monolayers one can obtain multiple meta-stable interlayer stacking sequences that may allow for even larger electrical polarization.**

In crystalline van der Waals (vdW) multilayers, one can distinguish between crystal polytypes, i.e., polymorphs with structures that differ only along the out-of-plane direction[12]. In a two-dimensional (2D) vdW polytype, each successive layer is identical, but can stack in different metastable configurations that modify overall crystal properties[a]. For example, parallel stacking of diatomic hexagonal bilayers has been recently shown to result in permanent out-of-plane polarization, $P_z$, which is absent in the natural antiparallel stacking configuration[8–10]. Interestingly, it was possible to switch between the polar configurations by sliding the partial dislocation that separates them, resulting in a phenomena called "interfacial ferroelectricity"[9,13,14]. Moreover, distinct multilayer polytypes exhibit cumulative polarization steps, known as "ladder ferroelectrics"[11,15]. This unique opportunity to switch locally between many adjacent diatomic polytypes provides a novel platform to explore their basic quasi-particle response on the one hand, and to exploit their polarization for storage and information processing technologies on the other hand.

---

[a] Here we use the term polytype in its broad sense[12], which includes metastable 2D arrangements.

Polarization in the above-discussed gapped diatomic polytypes emerges from an inherent charge transfer between atoms of different types. This is obviously lacking in graphene-layered polytypes - a semi-metallic system of ever-growing interest[16–25], where the constituent layers contain a single atomic species, and are centrosymmetric (as opposed to polar elemental crystals of Tellurium and Bismuth reported recently[26,27]). Could polarization emerge in this case at all? And, if so, what would its nature be? To explore these questions, recall that in a crystal based on one type of atom, unit cell symmetry and nuclear positions within the unit cell determine all crystal properties. Specifically, when graphene is stacked into a bilayer structure, the interface avoids the high-energy fully eclipsed (AA) configuration, adopting instead the optimal (AB) stacking mode, where only half of the atoms reside atop each other[28]. By introducing additional layers, multiple stacking configurations, distinguished by the relative lateral shift between the various layers[28–32], become accessible (see Fig. 1a,b). For example, in the case of three layers, the crystal can stabilize two high-symmetry polytypes, denoted by ABC or ABA, that preserve inversion [x→-x, y→-y, z→-z] or mirror (z→-z) symmetry, respectively. In a four-layer stack, the number of polytypes grows to three: (i) the Bernal (B); (ii) the rhombohedral (R) polytypes, both of which break mirror symmetry but preserve inversion symmetry; and (iii) the ABCB (denoted as P) polytype, which breaks both symmetries, allowing for permanent polarization.

Generally, for a stack of an even number $N$ of layers we find $2^{N-3} + 2^{(N-4)/2}$ different polytypes possible, all breaking mirror symmetry, out of which only $2^{(N-2)/2}$ possess inversion symmetry and are hence necessarily non-polar. For an odd number of layers in the stack, we find $2^{N-3} + 2^{(N-3)/2}$ polytypes, out of which $2^{(N-3)/2}$ are non-polar centrosymmetric stackings, and $2^{(N-3)/2}$ are mirror-symmetric configurations that may only present in-plane polarization. The rest of the stacking modes break both symmetries and may exhibit permanent $P_z$ and in-plane polarization (see detailed discussion in SI.1). For example, for 11 layers (N=11), there are 272 different stacking configurations, out of which 16 are non-polar, 16 may only exhibit in-plane polarization, and 240 may exhibit both out-of-plane and in-plane polarization (see Table S1).

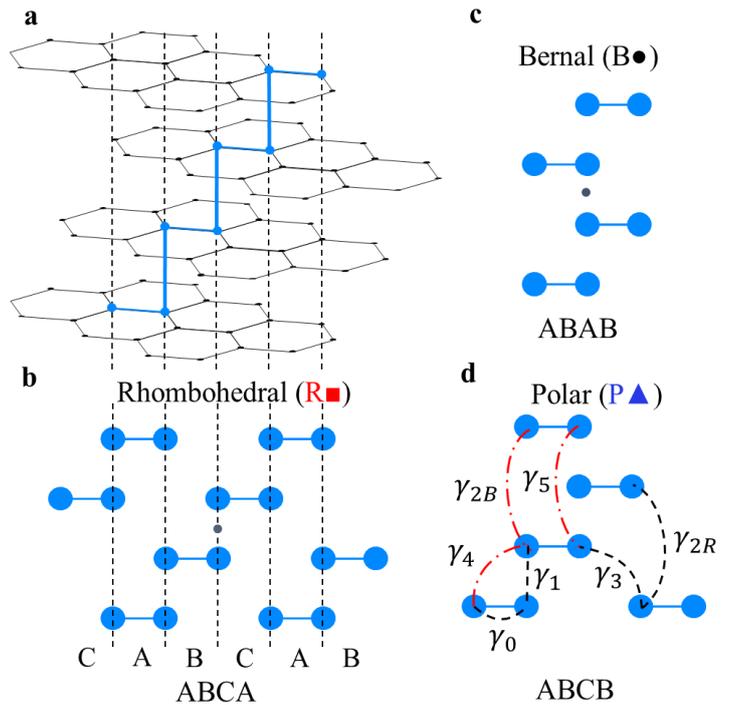

**Figure 1| Polytypes of four-layered graphene. a** illustration of the rhombohedral (R, ABCA) polytype, with co-oriented shifts between consecutive layers. **b**. Cross-section of the R polytype along the in-plane armchair orientation, showing a cyclic shift of carbon pairs between the A/B/C/A positions. **c**. Cross sectional illustration of the Bernal polytype (B, ABAB) with the inversion center marked by a black circle. **d.** Cross sectional illustration of the polar polytype (P, ABCB), lacking inversion and mirror symmetries. Various tight-binding model hopping amplitudes are indicated by red dash-dotted and black dashed lines, connecting atoms residing in the same or different sublattices, respectively.

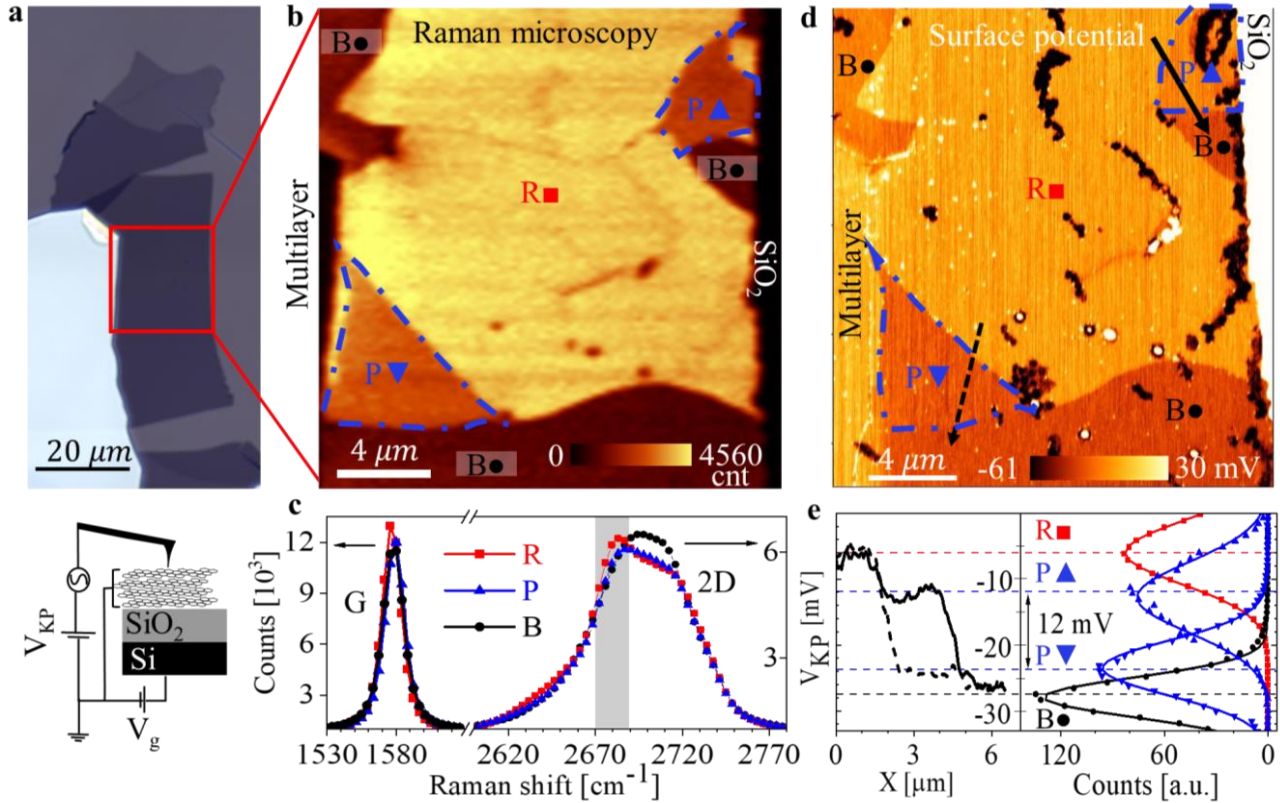

**Fig. 2| Direct observation of internal polarization in graphite. a.** Optical image of a typical tetra-layer flake on a Silicon/Silicon dioxide (90nm) substrate. The cross-sectional illustration at the bottom shows the Gate Voltage ($V_g$) and the Kelvin Probe potential ($V_{KP}$) circuits in the electric force microscopy setup. **b.** Integrated Raman Intensity map of the area marked in (**a**) with the color in each pixel representing the number of photons collected at the spectral range, shown as a grey bar in (**c**); The map shows three distinct brightness levels assigned, based on their Raman signatures, to the Rhombohedral (R■), Polar (P▲), and Bernal (B●) polytypes. **c.** Raman spectra collected from each polytype, showing distinct shifts and intensity variations in the G and the 2D peaks. **d.** Surface potential map obtained from the same area as in (b); B and R polytypes appear uniform in color and are separated by 17 mV. In contrast, the P polytype appears in two distinct colors. The black areas are typical contamination spots. **e.** Left: Potential cuts over R-P-B polytypes, along the solid and dashed lines marked in (d); Right: Normalized potential histograms from separate polytype regions and their fitting Gaussians (solid lines).

To explore the possibility of polarization in multi-layered all-carbon graphene stacks, we exfoliate natural graphite flakes on bare silicon oxide ($SiO_2$) surfaces. We determine the number of graphene layers in each flake from their optical contrast, as well as from their atomic force microscopy (AFM) measured thickness. The marked area in Figure 2a shows a typical tetra-layer graphene region with uniform optical and topography maps. However, the integrated Raman map of the same section (Fig. 2b), constructed by integrating the photon counts in the range 2674-2684 cm$^{-1}$ (see Fig.2c), reveals three distinct regions, which can be assigned to the above defined B, R, and P polytypes based on their known Raman signatures[25]. We further scan the same area by a Kelvin probe force microscope[33] to map the room temperature electric surface potential $V_{KP}$ (Fig. 2d). The contrast between polytypes is present also in the electrical measurement, reflecting variations in the work function of the different polytypes[22,34,35]. Remarkably, the two P polytype regions that appear with the same color in the Raman map, exhibit different colors in the $V_{KP}$ map, clearly demonstrating their opposite internal polarization $P_\uparrow$, $P_\downarrow$. Two line cuts across the polytype regions (solid and dashed black arrows) are presented in Fig. 2e, demonstrating distinct work function steps between the uniform values corresponding to individual polytypes. To quantify these variations, we consider potential histograms as a function of the number of pixels per potential value, plotted on the right-hand side of Fig. 2e. Repeating this analysis for many different samples (SI. 3.1) yields ($V_{KP}(P_\uparrow)$ - $V_{KP}(P_\downarrow)$)/2 = 6±1mV, $V_{KP}(R)$ - $V_{KP}(B)$= 19±1mV, $V_{KP}(R)$ - $V_{KP}(P_\uparrow)$ = 5±1mV. These numbers are independent of polytype dimensions, the AFM tool, the environment (ambient vs. inert), and the tip type. We note that the average map potential may vary by as much as ~200mV in different experiments, yet the potential steps between the polytypes remain unaltered.

The permanent room temperature polarization in a graphitic (seemingly semi-metallic) polytype calls for a careful examination of its doping dependence[11,36,37]. To this end, we biased the flakes relative to the bottom Si substrate, thereby forming a planar capacitor that controls the total charge carrier density of the flake[11] (Fig. 2a). We then re-mapped the surface potential for several fixed gate voltages, $V_g$, and extracted the potential steps between the different polytypes as a function of the planar charge density $n$. As expected, the new maps show ± 10V variations in $V_{KP}$ outside the graphite flakes, corresponding to the applied gate voltage, whereas above the flake $V_{KP}$ is altered by merely ±20mV (see Fig. S2)[11]. This corresponds to charge accumulation on the flake, with n =± 2.3×10$^{12}$ e/cm$^2$, which screens the bottom gate potential (SI 3.2). Figure 3a presents the gate and density evolution of the potential step between opposite polar domains ($V_{KP}(P_\uparrow)$ - $V_{KP}(P_\downarrow)$)/2. Surprisingly, the potential difference and its corresponding internal polarization increase by ~ 13% upon hole doping to 10$^{12}$ holes/cm$^2$. Such an increase was not observed in layered diatomic structures or in any other system we are aware of, where the added charge distribution typically suppresses the internal polarization.

To rationalize the emergent polarization and explain its unusual dependence on doping, we conducted density functional theory (DFT) calculations (SI. 4.1). The overall $P_z$ is extracted from the difference between the laterally averaged electrostatic potential far below and above the 2D periodic stack[38], calculated via $\phi(z) = -\int_{-\infty}^{z}(z'-z)[\int dxdy \rho(x,y,z')]dz'$. As expected, the calculated potential is mirror symmetric for the B and R phases, where the polarization vanishes (Fig. 3b, black squares and red dots, respectively). However, for the four-layered P-polytype, a potential difference of ~ 6 mV appears at an electronic temperature of 300K (blue triangles), in good agreement with the measured ($V_{KP}(P_\uparrow)$ - $V_{KP}(P_\downarrow)$)/2 value (Fig 3a). Note the lower potential at the third layer plane of the polar phase (blue curve) versus the B and R phases, indicating a negative charge accumulation in this layer. Fig. 3b also shows a cross section of the charge redistribution along the armchair direction after subtracting the charge of the individual layers. The third layer exhibits a pronounced charge difference between the two sublattice sites, while the first and fourth layers show a weaker and opposite response. Next, we introduce effective doping in the calculation by including fractional nuclear charge "pseudoatoms" that induce excess free charge carriers without violating sample neutrality or distorting the underlying band structure (SI 4.1.2)[38,39]. The computed doping dependence of the polarization (red curve in Fig. 3a) is in overall good agreement with the experiment, showing a slightly larger slope of ~18% per 10$^{12}$ holes/cm$^2$.

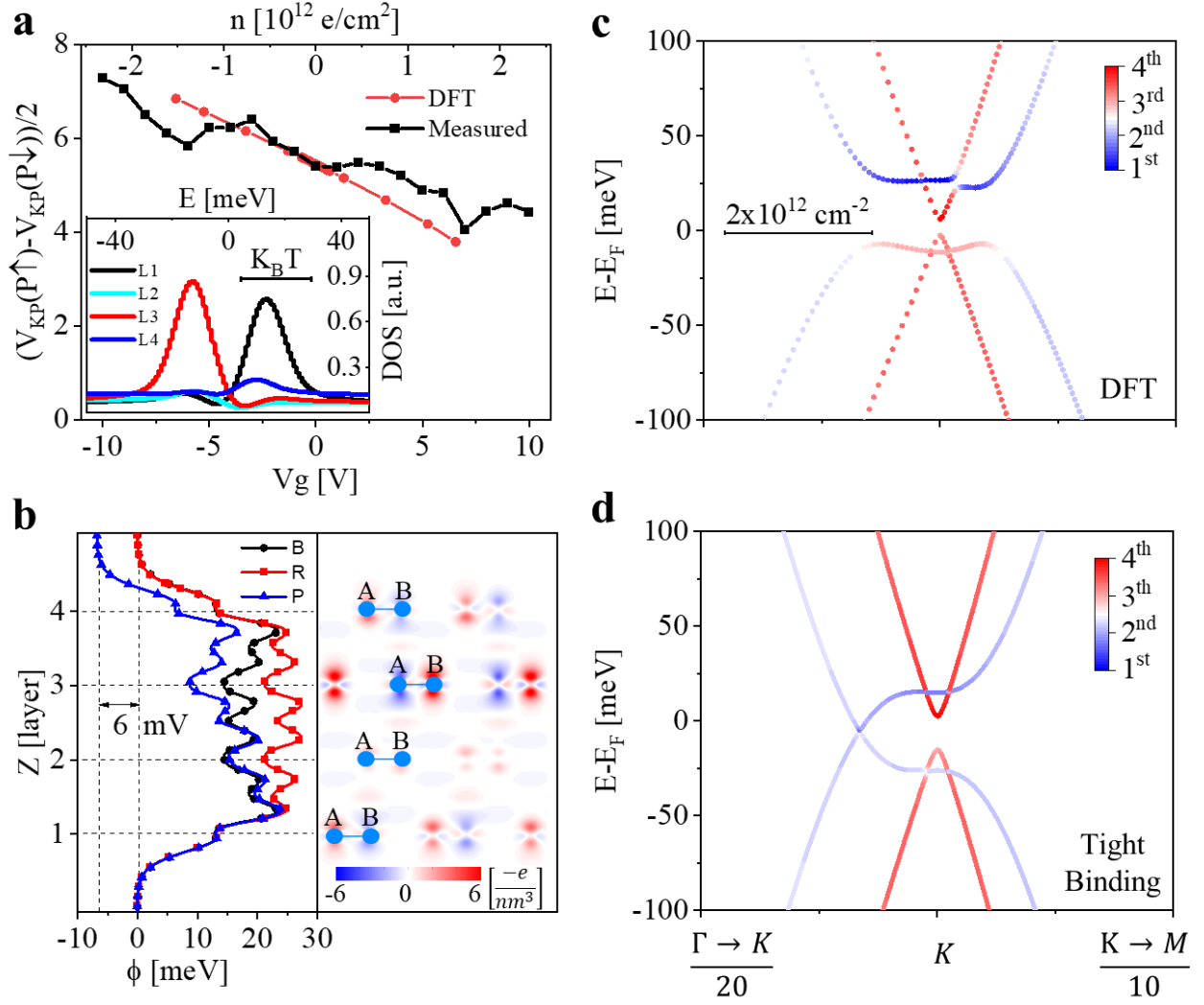

**Fig. 3| Dispersion and doping-dependent polarization of the P tetra-layer polytype. a.** The measured (black squares) and calculated (red circles) out-of-plane polarization ($V_{KP}(P\uparrow) - V_{KP}(P\downarrow))/2$ as a function of the external gate voltage (lower horizontal axis) and the corresponding charge density n (upper horizontal axis). Inset: DFT calculated DOS projected on each layer (L1 − L4) as a function of the energy. **b**. Left: The planar averaged electrostatic potential as a function of the vertical coordinate, $Z$, in the three graphitic polytypes. Right: Charge density redistribution map of a cross-section along the armchair direction of the P polytype. Red/blue colors correspond to the residual finite electron/hole density after subtracting the charge distribution of the individual monolayers. **c.** and **d.** DFT and self-consistent tight-binding calculated band structures, respectively, centered around the reciprocal K point. The color of the circular symbols marks the wave function projection of the crystal-momentum states on each layer. The scale bar indicates the momentum diameter required to occupy $2 \times 10^{12}$ cm$^{-2}$ states.

The origin of this doping dependence can be rationalized by examining the band structure at the edge of the Brillouin zone near the K point[28,30], where the Fermi level resides (Fig, 3c). Unlike the case of the B and R phases (Fig S6), a gap opening of ~9 mV appears between the top of the linear valence band and the bottom of the linear conduction band of the P phase, due to the broken inversion and mirror symmetries. Notably, contributing the most to the density of states (DOS) in the vicinity of the Fermi energy (see Fig. 3c) are the flat bands (separated by ~50mV) that are localized on the 1$^{st}$ (red) and 3$^{rd}$ (blue) layers. From the projected DOS (see inset of Fig. 3a) it becomes evident that swapping the gate potential from positive to negative around the Fermi energy depopulates the first and third layers. Because the width of the room-temperature Fermi-Dirac distribution ($2K_BT\sim50mV$) encompasses both peaks in the DOS (see inset of Fig. 3a), the two layers depopulate by approximately the same charge. However, the center of this extra charge distribution is shifted from the center of the tetralayer stack and therefore the polarization increases.

Further insights regarding the microscopic origins of the polarization can be obtained via a tight-binding (TB) model that includes a self-consistent electrostatic potential term. We consider two sites per layer with previously suggested hopping parameters and on-site energies[40] (S4.2). The hopping parameters are illustrated in Figure 1d, where we distinguish between coupling of sites on different sublattices (dashed black lines) and on the same sublattice (dash-dotted red lines). An important insight is that if the same-sublattice small amplitudes are neglected, polarization will vanish at charge neutrality due to particle-hole symmetry. Hence our experiment provides a very sensitive probe of them. Still, even when neglecting these terms, polarization would generally become nonzero upon gating. Accounting for these terms, the corresponding polarization will generally decrease with one sign of the doping, but still increase with the other. This behavior, which is indeed the experimentally observed one, then emerges as typical for graphene polytypes and is inherently different from that observed in polyatomic stacks. The resulting TB band structure is shown in Fig. 3d. Several general observations emerge (S4.2): (a) If the self-consistent electrostatic potential is omitted, the calculated polarization disagrees with the experimental and the DFT one in both magnitude and sign. (b) Allowing for the intra-layer potential differences is crucial for obtaining good agreement with the DFT results; (c) The electron-hole band asymmetry is generally more pronounced in the tight-binding dispersion than in the DFT one.

To conclude, we have demonstrated that judiciously chosen polytype architectures of non-polar building blocks may exhibit electric polarization. This was achieved in stacks of non-polar graphene monolayers which breaks inversion and mirror symmetries. We found the room temperature polarization of the graphene multilayer stack to decrease with electron doping but counterintuitively increase with hole doping. We attribute this behavior to gating-induced interlayer charge transfer, as reflected by the calculated band structure of the non-centrosymmetrically stacked system. This paves the way to control of the out of plane polarization of graphene polytypes via an external gate potential, despite the semi-metallicity and the non-polar nature of the individual graphene sheets. Furthermore, the general nature of the discovered effect calls for exploring stacking-induced polarization effects in the numerous asymmetric polytypes of graphene and other elemental vdW materials.

## **Acknowledgements**

We thank N. Ravid and I. Malker for laboratory support. VF acknowledges support from EC-FET Core 3 European Graphene Flagship Project, EPSRC grants EP/S030719/1 and EP/V007033/1, and the Lloyd Register Foundation Nanotechnology Grant. M.G. has been supported by the Israel Science Foundation and the Directorate for Defense Research and Development grant no. 3427/21 and by the US-Israel Binational Science Foundation grant no. 2020072. L.K. thanks the Aryeh and Mintzi Katzman Professorial Chair and the Helen and Martin Kimmel Award for Innovative Investigation. M.U. acknowledges the financial support of the Israel Science Foundation, grant no. 1141/18, and the binational programme of the National Science Foundation of China and Israel Science Foundation, grant no. 3191/19. O.H. is grateful for the generous financial support of the Israel Science Foundation under grant no. 1586/17, The Ministry of Science and Technology of Israel (project no. 3–16244), the Heineman Chair in Physical Chemistry, and the Naomi Foundation for generous financial support from the the 2017 Kadar Award. M.B.S. acknowledges funding by the European Research Council under the European Union's Horizon 2020 research and innovation programme (grant agreement no. 852925), and the Israel Science Foundation under grant nos. 391/22 and 3623/21. O.H. and M.B.S. acknowledge the Centre for Nanoscience and Nanotechnology of Tel Aviv University.

# Spontaneous Electric Polarization in Graphene Polytypes

# Supplementary Information.

Table of Contents

# S1. Crystalline Polytypes of Multilayer Graphene

The term polytype refers to any of a number of forms of a crystalline substance, the arrangement of which differs only along one of the unit-cell dimensions. In bilayer graphene, given that the fully eclipsed high-symmetry AA stacking configuration is unstable, there is one polytype with two equivalent stacking configurations, namely AB and BA. The same logic follows for $N$-layered graphitic stacks, where only a part of the possible $2^{N-1}$ high-symmetry stacking configurations accessible through layer sliding exhibit distinct dispersion relations.

To count the number of distinguishable crystals and to identify the polarization properties of each polytype, it is instructive to classify them according to their inversion (I) (x→-x, y→-y, z→-z) and mirror (M) (z→-z) symmetries. If a point in space exists for which (I) results in the same structure, the system is non-polar. If I is broken, permanent in-plane polarization may emerge for M-symmetric structures. Additional out-of-plane polarization may be found if the M symmetry is also broken.

We note that for AB/BA stacking:
a. Polytypes with no I and no M symmetry have four equivalent stacking configurations, obtained from one polytype by applying I, M, or I and M operations.
b. Polytypes with either I or M have two degenerate stacking configurations, obtained from one polytype by applying the I or M operation.
c. There are no polytypes with both I and M.

For example, in tri-layers $N=3$ and there are $2^{3-1}=4$ possible stacking configurations, but only two distinct polytypes: B (M symmetric) and R (I symmetric). In tetra layers, $N=4$, the B and R polytypes are I symmetric (and therefore each is counted twice according to the above rules), while the P polytype breaks both I and M and hence is counted four times. Subtracting the over-counted configuration degeneracies out of the total of $2^{4-1}=8$ configurations, we find that there remain only 8-1-1-3=3 unique polytypes: B, R, and P.

We further note that if the number of layers $N$ is even, then no polytypes are M symmetric due to the shift at the middle interfacial plane, and $2^{(N-2)/2}$ polytypes are I symmetric. To see this, consider fixing the inversion point between the eclipsed atoms of the two middle layers (see Fig. 1b,c in the main text) and only counting the number of stacking options above the middle plane. The stacking configurations below the middle plane are then fixed to conserve I. Hence, they do not add new configurations to the count. If the number of layers $N$ is odd, then $2^{(N-3)/2}$ polytypes are M symmetric, and $2^{(N-3)/2}$ polytypes are I symmetric. To see this, consider fixing the inversion point between the two atoms of the middle layer and counting the number of stacking options above it only. The stacking configuration below is then fixed to either conserve I or conserve M and hence does not introduce new polytypes to the count.

To summarize, for even/odd $N$:
a. $2^{(N-2)/2}$ / $2^{(N-3)/2}$ polytypes are I symmetric (and must break M symmetry). These polytypes are not polar.
b. 0 / $2^{(N-3)/2}$ polytypes break I symmetry but conserve M symmetry. These polytypes are polar in plane only.
c. All remaining polytypes, the number of which is $\frac{1}{4}(2^{(N-1)} - 2\times2^{(N-2)/2})$ / $\frac{1}{4}(2^{(N-1)} - 2\times2\times2^{(N-3)/2})$, break both I and M symmetries and can be polar both in and out of the plane.

Table 1 summarizes the above combinatorial count for $N$ =2 to 20. For example, it shows that there are 136 distinct polytypes for ten layers and 131,328 distinct polytypes for 20 layers. As also shown in the Table, most of these configurations exhibit broken I and M symmetry and therefore support out-of-plane polarization.

| N | No P (I Mz) | IP P (I̸ Mz) | OOP+IP (I̸ M̸z) | Total |
|---|---|---|---|---|
| 2 | 1 | 0 | 0 | 1 |
| 3 | 1 | 1 | 0 | 2 |
| 4 | 2 | 0 | 1 | 3 |
| 5 | 2 | 2 | 2 | 6 |
| 6 | 4 | 0 | 6 | 10 |
| 7 | 4 | 4 | 12 | 20 |
| 8 | 8 | 0 | 28 | 36 |
| 9 | 8 | 8 | 56 | 72 |
| 10 | 16 | 0 | 120 | 136 |
| 11 | 16 | 16 | 240 | 272 |
| 12 | 32 | 0 | 496 | 528 |
| 13 | 32 | 32 | 992 | 1056 |
| 14 | 64 | 0 | 2016 | 2080 |
| 15 | 64 | 64 | 4032 | 4160 |
| 16 | 128 | 0 | 8128 | 8256 |
| 17 | 128 | 128 | 16256 | 16512 |
| 18 | 256 | 0 | 32640 | 32896 |
| 19 | 256 | 256 | 65280 | 65792 |
| 20 | 512 | 0 | 130816 | 131328 |

Table S1 Combinatorial count of graphene polytypes obeying I Symmetry (no polarization), vertical M symmetry (in-plane polarization only), or no symmetry (in- and out-of-plane polarization)

## S2. Measurements Details

### S2.1 Raman microscopy

Raman measurements are performed in a commercial WITEC alpha300 Apyron confocal microscope equipped with a UHTS 300 mm focal length spectrometer. We use a 532 nm laser focused down to a spot size of ~300 nm, with a 300 lines/mm grating corresponding to a spectral range of ~3800 cm$^{-1}$ and a spectral resolution of ~ 4 cm$^{-1}$. The G (~1580 cm$^{-1}$) and 2D (~2700 cm$^{-1}$) peaks are taken in a single scan (Fig. 2b). The raster scans are obtained by using a step size of 200 nm, an average power of ~4 mW, and integration times shorter than 1 sec to avoid laser-induced heating.
To construct the Raman maps (Fig 2, Fig S1) we integrate the photon counts in a specific range that optimizes the contrast between polytypes. The filter range of the maps is shown in panel b and c next to it. We note that no damage or boundary wall movements are observed during the scans.

## S2.2 AFM measurements

The topography and KPFM measurements are performed using two separate microscopes: (i) PARK NX-10 in ambient conditions, and (ii) PARK NX-HIVAC in a nitrogen gas environment at a pressure of 100 mbar. We use a PPP-EFM metal-coated tip with a mechanical resonance frequency of ~75 kHz and a spring constant of 3 N/m. The cantilever amplitude at the non-contact modes ranged between 10 and 30 nm. The topography and KPFM signals are obtained separately using a two-pass measurement. The first pass records the topography. In the second pass, the KPFM (DC) potential is recorded by lifting the tip an extra 7 nm and repeating the topography profile of the first pass under a bias voltage of 1.5V AC: 1.5-3 kHz. The side-band KPFM mode is used to obtain a better-defined local potential.

We find variations of ~ 200 mV in the average flake potential across different samples, microscopes, and tips. We attribute these variations to different environmental conditions, the charging of the substrate, and tip contaminations. Our conclusions, however, rely only on local surface potential differences between adjacent polytypes, which we find to be independent of all the above-mentioned factors.

We further note that there is a slight potential gradient within each map (a nearly linear variation of a couple of mV over a 10 μm scale), which we attribute to non-local interactions with the macroscopic cantilever. This low-frequency variation is eliminated by standard image processing tools (Gwyddion software).

While clearly detected in line cuts (Fig. 2e, Fig S1), the surface potential above each polytype is quantified more accurately by considering the potential of all the relevant pixels in each polytype region. Indeed, plotting the potential histogram of the adjacent polytypes (after masking out regions of local contaminations) allows us to detect the mean surface potential of a polytype to within ~1 mV.

# S3. Additional experimental results

## S3.1 Polarization measurements in adjacent polytypes

In total, we identified and characterized nine different tetra-layer graphene flakes, each containing one or more polar polytypes. We note that only a small fraction of the exfoliated flakes, ~ 5%, include a polar polytype. Altogether, these nine flakes included 19 polar polytypes (ten domains pointing up and nine pointing down). Within each flake the different polytypes P, R and B exhibit consistent surface potential differences: $(V_{KP}(P_\uparrow) - V_{KP}(P_\downarrow))/2 = 6\pm1$mV, $V_{KP}(R) - V_{KP}(B) = 19\pm1$mV, $V_{KP}(R) - V_{KP}(P_\uparrow) = 5\pm1$mV, as discussed in the main text.

The polytypes are distinguished by filtering the Raman counts near the 2D or G peaks, as shown in Fig. S1; Interestingly, the region marked by a solid line frame in the figure, which is presented with high resolution in panel (c), contains the two oppositely polarized domains separated by a single domain wall.

**Figure S1| Additional samples demonstrating polar domains of opposite polarization. a.** Optical image, Raman map, and a surface potential map of a selected tetralayer flake. The Raman map is constructed using a filter near the 2D peak as shown in panel (c). **b/c.** Zoom-in maps on the region framed with a dashed/solid line in both the Raman and surface potential maps of (a). The zoomed-in Raman map is constructed using a filter near the G or 2D peak, noted by a gray bar in the panels. The two domains of the polar polytype (blue arrows) appear in the same color in the Raman maps, but show distinct colors in the surface potential maps. The normalized histogram of the number of pixels per a given surface potential, for each of the polytypes, is fitted with a gaussian curve. Curves of opposite P polytypes are separated by 13 mV in both frames. A further potential line-cut is shown in (c) across the adjacent oppositely polarized domains.

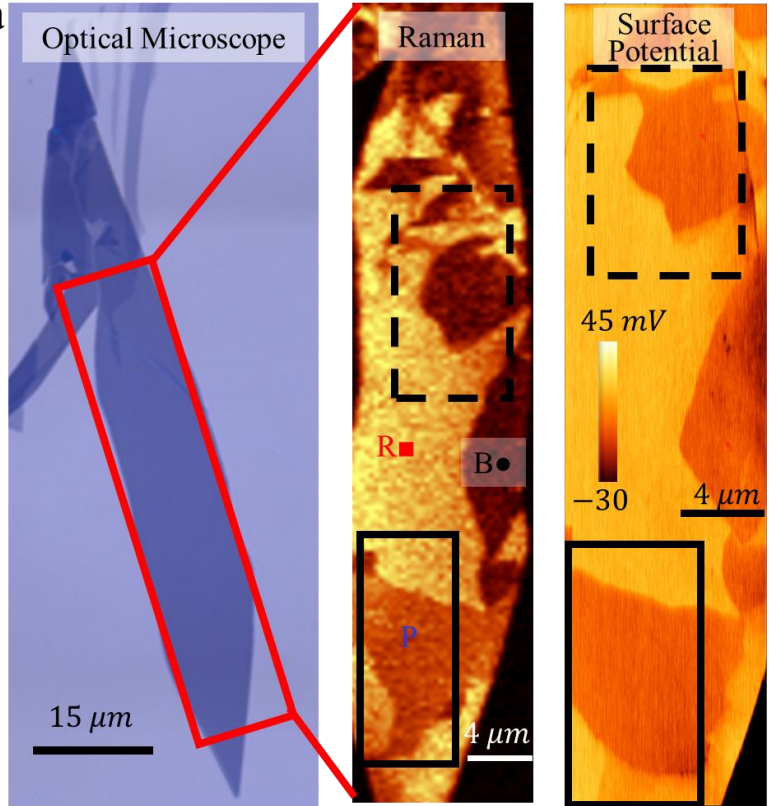

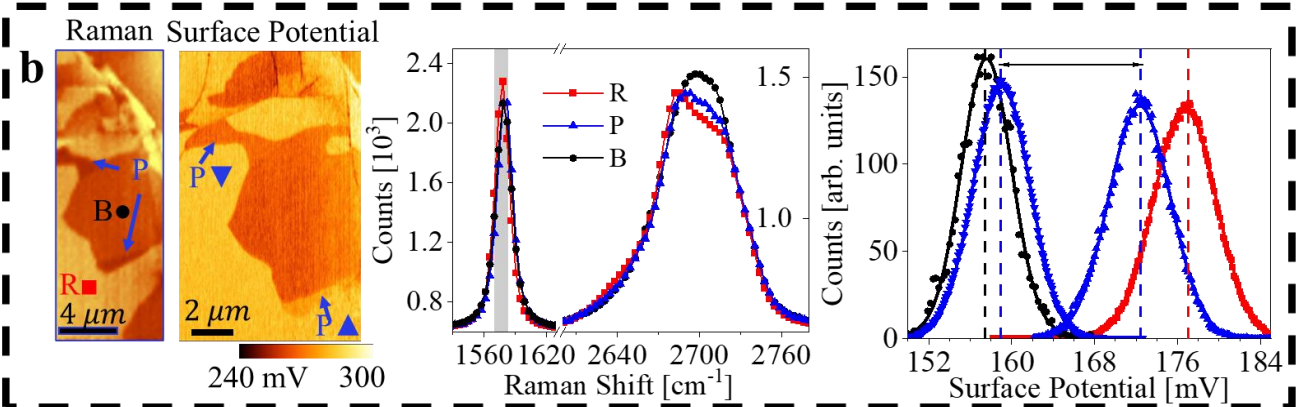

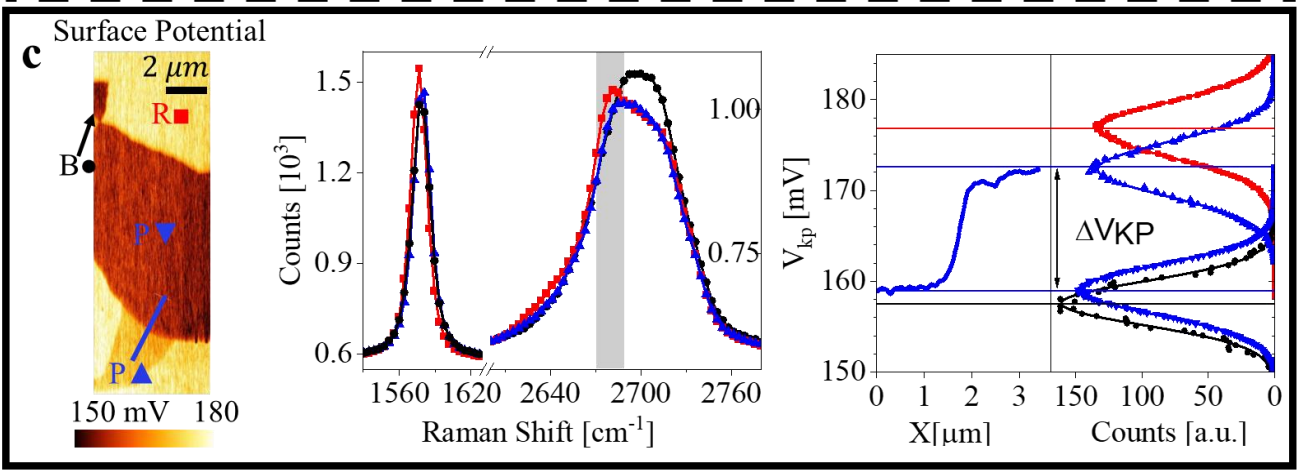

## S3.2 Polarization measurements as a function of doping

In the doping-dependent measurements, we apply a gate voltage between the bottom silicon electrode and the graphene, as shown in Figure 2 of the main text. We estimate the charge density from the planar plate capacitance as $n/V_g = \frac{\epsilon_0 \cdot \epsilon_K}{e \cdot d} = 2.3 \times 10^{12}\ cm^{-2} V^{-1}$. This common procedure is confirmed on many other flakes by transport measurements[1], where we find an intrinsic doping level (at zero $V_g$) that is smaller than $3\times 10^{11}\ cm^{-2}$. As expected, the potential atop the silicon surface changes according to the applied voltage (in the range of ±10 V), while above the tetra-layer graphenes, $V_{KP}$ changes by ±15 mV over the same $V_g$ range (Figure S2). We attribute the latter potential shift observed in all polytypes to the density-dependent Fermi energy and its impact on the work function[2]. The polarization magnitude as a function of $V_g$ and $n$ is extracted from the potential difference between oppositely polarized domains (of otherwise the same material and work function).

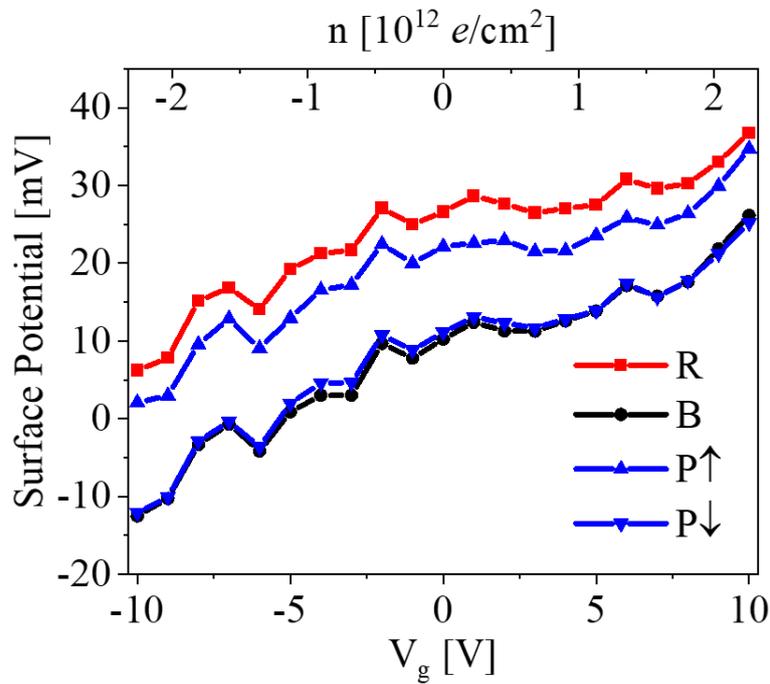

**Figure S2|** Surface potential variations as a function of gate voltage for the rhombohedral (R), Bernal (B), and two oppositely polarized polar domains (P) tetra-layer polytypes at room temperature.

# S4 Theoretical modeling

## S4.1 DFT

### S4.1.1 Calculation of potential profiles in graphene polytypes

Computational details

The laterally integrated electrostatic potential profiles along the normal direction of three polytypes of four-layered graphene are shown in Fig. 3b of the main text. To obtain the profiles, we used the Perdew-Burke-Ernzerhof (PBE) generalized-gradient exchange-correlation density functional approximation[3], augmented by the Grimme-D3 dispersion correction using Becke-Johnson (BJ) damping[4], as implemented in the Vienna Ab-initio Simulation Package (VASP)[5]. A plane wave energy cutoff of 1000 eV and a k-point mesh of $90 \times 90 \times 1$ were used, with a vertical vacuum size of 10 nm to avoid interactions between adjacent images. The core electrons of the carbon atoms were treated via the projector augmented wave (PAW) approach.

The polytypes have been constructed by stacking four relaxed monolayers in the B, R, and P configurations, and further relaxing the entire stack using the conjugated gradients algorithm with a force threshold of $10^{-3}$ eV/Å. Single-point electron density calculations were then performed on the relaxed structure using a Gaussian smearing of 25.8 meV, to enhance the convergence of the self-consistent cycle.

Consistency and convergence tests

To evaluate the vertical polarization, a dipole moment correction was employed[6]. For validation purposes, double supercell calculations were also performed (see Fig. S3a), where the supercell consists of two opposing mirror images of each graphene stack with a 6 nm inter-image vacuum region. Fig. S3b shows that the dipole correction and double supercell methods yield nearly identical potential profiles with an overall electrostatic potential drop of 6.7 meV.

Convergence tests of the VASP calculations (Fig. S4) indicate that our choice of parameters leads to electrostatic potential differences convergence to within 0.2, 0.001, and 0.001 meV with respect to the number of k-points, energy cut-off, and vacuum size, respectively. The corresponding binding energies convergence values are 0.0003, 0.0006, and 0.003 meV/atom. Note that a proper description of the potential profile requires a relatively dense mesh of 721 k-points.

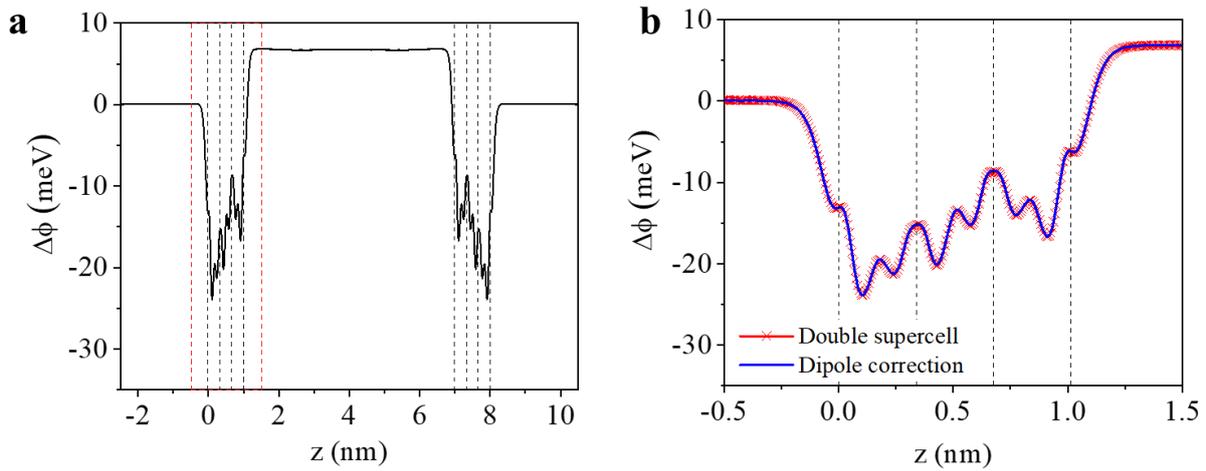

Figure S3. **Potential profiles for a P polytype tetra-layered graphene stack.** (a) Difference between the laterally averaged potential profile obtained for the four-layered graphene stack double supercell and that obtained from a superposition of the corresponding non-interacting monolayers. (b) Comparison between the potential profiles obtained using the dipole correction (blue), and the double supercell method (red), in the dashed-red rectangular region denoted in panel (a). The black dashed lines represent the vertical locations of the four layers. The origin of the horizontal axis is set to the bottom layer.

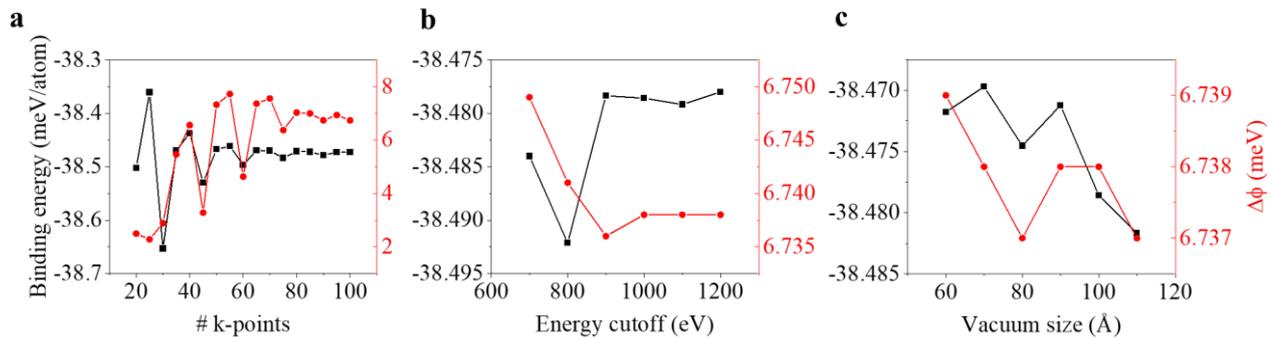

Figure S4. **VASP calculations convergence tests.** Convergence tests of the binding energy (black curve, left vertical axes) and electrostatic potential difference across the entire structure (red curve, right vertical axes) for the P polytype of a four-layered graphene stack with respect to: (a) number of k-points; (b) energy cutoff, and (c) vacuum size.

Charge redistribution analysis

The origin of the observed vertical polarization can be traced back to charge redistribution in the non-centrosymmetric polytypes. This is demonstrated in Fig. S5, where charge redistribution and potential difference maps for a 2d cross-sectional cut along the armchair direction ((110) surface) of three polytypes is presented. The variation map is obtained from the difference between the charge density of the four-layered stack and the super-imposed densities of the corresponding isolated monolayers. The charge density difference map of the P polytype is clearly non-centrosymmetric (Fig. S5a), demonstrating a relatively pronounced in-plane dipolar pattern on the third layer with opposite weaker patterns on the first and fourth layers. Similar but centrosymmetric patterns are found for the R (Fig.

S5b) and B (Fig. S5c) polytypes, where the dipolar structure of the former is mainly localized at the edge layers and that of the latter is more pronounced on the middle layers. The corresponding electrostatic potential differences also manifest the broken symmetry of the P polytype (Fig. S5d), and the centrosymmetry of R (Fig. S5e) and B (Fig. S5f) polytypes.

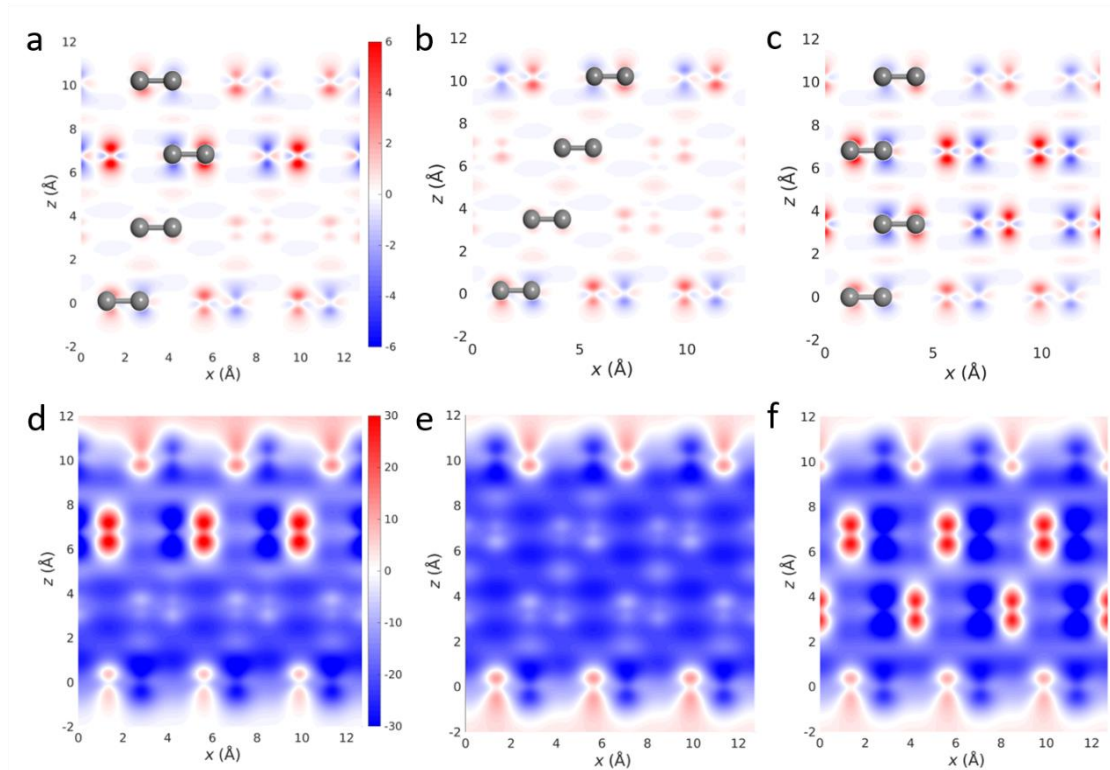

Fig. S5. **Charge redistribution and potential difference maps**. A two-dimensional cross section through the (a)-(c) charge density differences (with respect to the individual monolayers) and (d)-(f) potential difference along the (110) crystallographic plane of the P (a, d), R (b, e) and B (c, f) four-layered graphene polytypes. The scale-bar units for the charge density and potential difference maps are $e/nm^3$ and meV, respectively. The gray spheres appearing in the upper panels represent the positions of the carbon atoms.

Band structures of the B and R phases
In Fig. 3 of the main text we provide the calculated band-structure of the P polytype to analyze the origin of the observed vertical polarization. For completeness, we present in Fig. S6 the corresponding band structures of the B and R polytypes calculated at the level of theory described above.

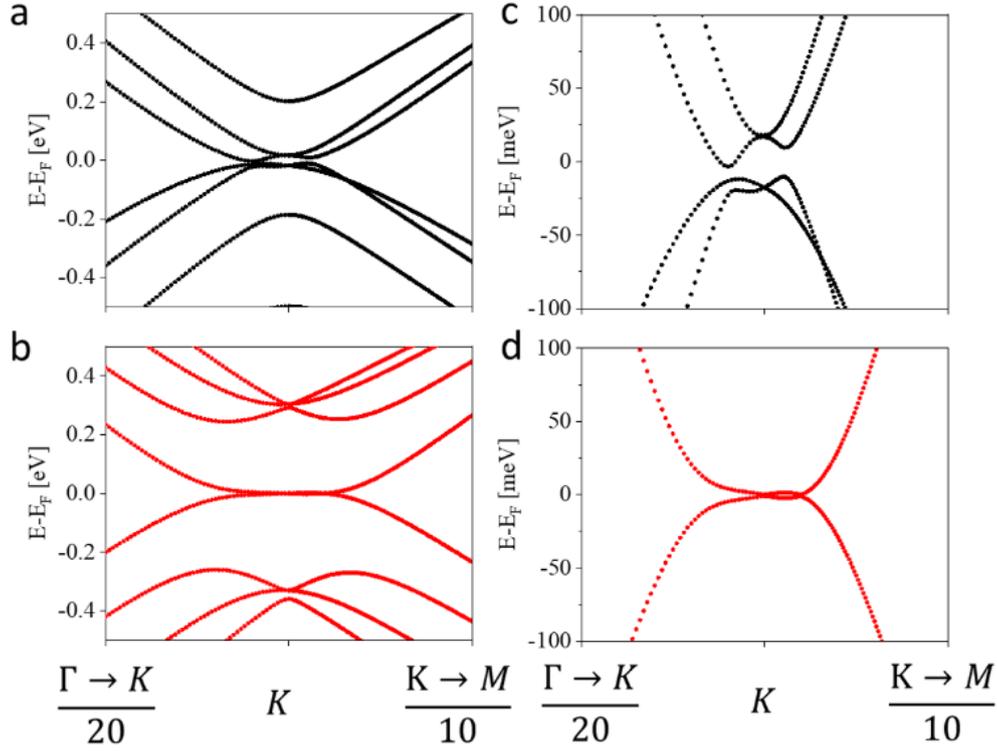

Fig. S6. **Band structures**. Calculated band structures of the (a) B, and (b) R, four-layered graphene polytypes. A zoom-in on the energy range near the Fermi energy is given in panels (c) and (d), respectively.

S4.1.2 Calculation of doping dependence of polarization in P polytype graphene

Doping calculations of P polytype four-layered graphene were performed using the fractional nuclear charge pseudoatom approach[7], allowing for simulating doping densities in the experimentally relevant range. To this end, we use pseudopotentials (PPs) generated for atoms with fractional nuclear charge. These calculations were performed using the open-source plane-wave Quantum Espresso package[8] (instead of VASP that was used to perform the calculations described in section S2), allowing us to construct appropriate PPs. We first generated Rappe-Rabe-Kaxiras-Joannopoulos (RRKJ)[9] PPs using the ld1.x program[8][10], while setting the nuclear charge of the carbon pseudoatom to $Z = 6 \pm |\varepsilon|$, the original charge of the neutral element plus a small fractional charge $|\varepsilon|$. The valence electronic charge was changed accordingly to maintain neutrality of the unit-cell, with an electron configuration given by $[He]2s^2 2p^{2\pm\varepsilon}$. A set of PPs were generated by setting $\varepsilon = 10^{-9}, 10^{-8}, \ldots, 10^{-4}$ for all C atoms in the system, corresponding to doping densities of $\Delta n_{2D} = 1.5 \times 10^7, 1.5 \times 10^8, \ldots, 1.5 \times 10^{12}$ cm$^{-2}$, respectively.

Single point calculations were performed using the generated PPs to obtain the electron density and the corresponding electrostatic potential profiles. To this end, we employed the PBE generalized-gradient density functional approximation[3] and the Grimme-D3 dispersion correction with BJ damping[4], as implemented in Quantum Espresso. A plane wave energy cutoff of 60 Ry (816.34 eV) was used with a k-mesh of $90 \times 90 \times 1$, and a vertical vacuum size of 10 nm was set to avoid interactions between adjacent bilayer images. Fermi-Dirac smearing with an effective temperature of ~300 K was used to enhance the convergence of the self-consistent cycle. To obtain converged electrostatic potential profiles, a dipole correction was used[6]. Similar to section S4.1.1, we also compared the results against calculations using a double supercell, confirming the consistency of our results (see Fig. S7).

As in the procedure discussed in section S4.1.1, undoped P polytype graphene was first constructed and optimized using the Quantum Espresso package, yielding an electrostatic potential drop of 5.4 meV. This value is somewhat smaller than the value of 6.7 meV obtained using VASP (see section S4.1.1), likely reflecting the different pseudopotentials used and other differences in numerical settings. Note, however, that the difference is of the order of 1 meV, which would normally have been considered as very small and only somewhat stands out because the overall dipole is small.

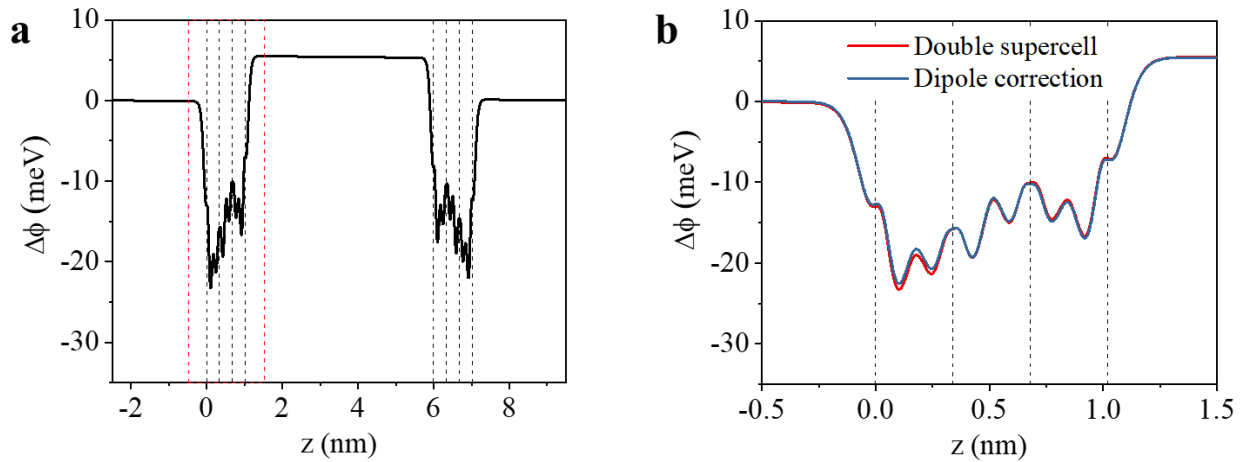

Figure S7. **Quantum Espresso potential profiles for a P polytype tetra-layered graphene stack.** (a) Difference between the laterally averaged potential profile obtained for the four-layered graphene stack double supercell and that obtained from a superposition of the corresponding non-interacting monolayers. (b) Comparison between the potential profiles obtained using the dipole correction (blue), and the double supercell method(red), in the dashed-red rectangular region denoted in panel (a). The black dashed lines represent the vertical locations of the four layers. The origin of the horizontal axis is set to the bottom layer.

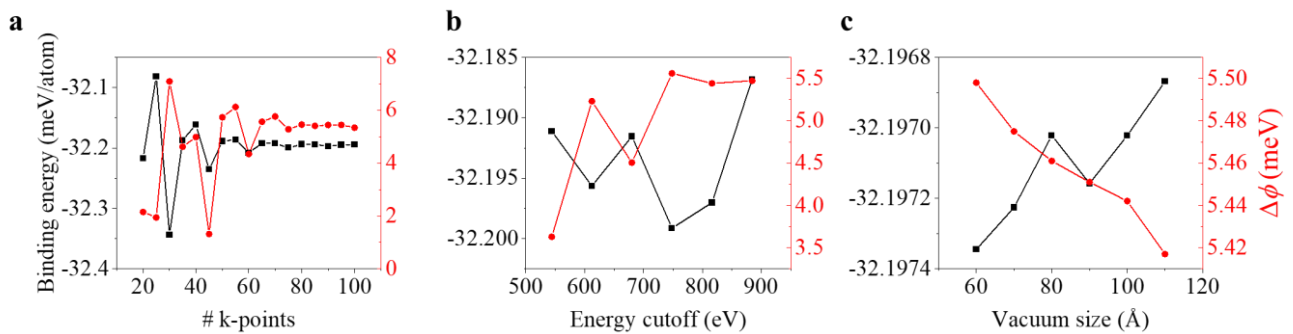

Figure S8. **Quantum Espresso calculations convergence tests.** Convergence tests of the binding energy (black curve, left vertical axes) and electrostatic potential difference (red curve, right vertical axes) for the P polytype of a four-layered graphene stack with respect to: (a) number of k-points; (b) energy cutoff, and (c) vacuum size.

Convergence tests of the Quantum Espresso calculations (Fig. S8) indicate that our choice of parameters leads to electrostatic potential differences convergence to within 0.002, 0.03, and 0.025 meV with respect to the number of k-points, energy cut-off, and vacuum size, respectively. The corresponding binding energies convergence values are 0.002, 0.01, and 0.0001 meV/atom.

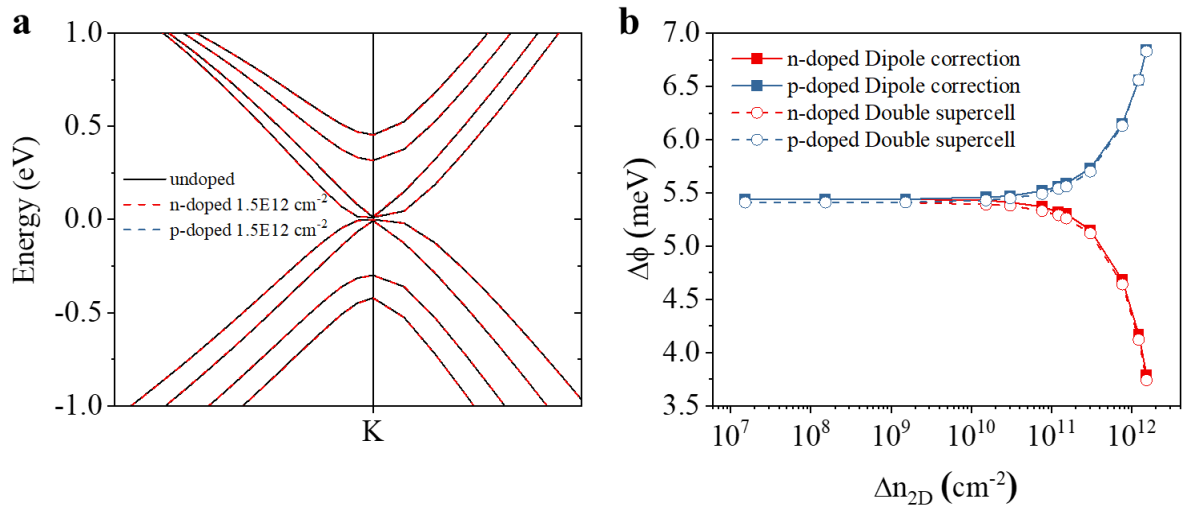

Figure S9. **Effect of doping on the band structure and potential drop.** (a) The band structures of undoped (black) and $1.5 \times 10^{12}$ cm$^{-2}$ $n$-(red) and $p$-(blue) doped P polytype graphene. The origin of the vertical axis is set to the topmost K-point valence band energy. (b) Potential drop as a function of electron ($n$, red) and hole ($p$, blue) doping densities, calculated using the dipole correction (solid curves, squares) and the double supercell (dashed curves, circles) methods.

We note that the fractional nuclear charge pseudoatom doping approach[9] adopted in this study remains valid as long as variations in the calculated band-structure, induced by the nuclear pseudo charging, are negligible. To confirm that our calculations satisfy this condition, we compare the band-structures of the undoped and doped cases up to the highest doping density considered (see Fig. S9a). Our results clearly demonstrate merely minor deviations of the band-structures of the doped systems from those of the undoped counterparts. The energy difference between the topmost K and Γ valence band points for the doped and undoped systems ($< 0.1$ meV) is sufficiently small to be neglected.

Upon doping, the polarization remains mostly unaffected up to a threshold value of ~$10^{10}$ cm$^{-2}$, above which a polarization decrease (increase) is clearly seen for n-(p-)doping (see Fig. S9b). Notably, the dipole correction and the double supercell approaches generate nearly identical doping induced polarization profiles, indicating the reliability of our results.

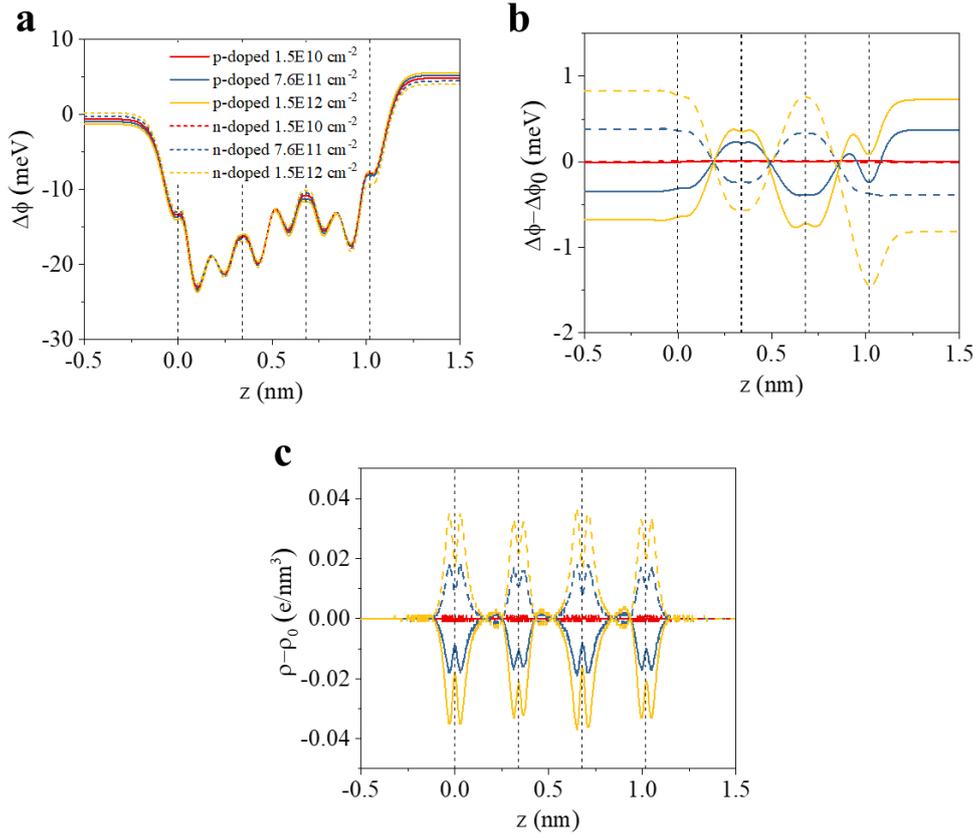

Figure S10. **Doping-induced charge and potential variations.** (a) Laterally averaged vertical potential profiles obtained at p- (solid lines) and n- (dashed lines) doping densities of $1.5 \times 10^{10} \text{cm}^{-2}$ (red), $7.6 \times 10^{11} \text{cm}^{-2}$ (blue), and $1.5 \times 10^{12} \text{cm}^{-2}$ (yellow) after subtracting the superposition of potential profiles of the corresponding monolayers. (b) Doping induced variation of the vertical potential profiles. (c) Doping induced variations of the laterally averaged charge density.

In the main text, we explore the origin of the doping induced polarization variations in terms of the projection of the electronic density of states on the different layers. For completeness, we discuss below how the corresponding potential profile variations are manifested in the charge density redistribution. Fig. S10a shows the laterally-averaged vertical potential profiles (after subtracting the potential profiles of the corresponding non-interacting monolayers) for doping densities near and above the threshold value, demonstrating a somewhat different response of the various layers. A clearer picture is obtained when subtracting the profile of the undoped system (Fig. S10b) showing a strong positive response of the first and third layers, a strong negative response of the fourth layer and a weaker negative response of the second layer. This asymmetric response is also manifested in the laterally averaged charge density difference profiles (Fig. S10c), demonstrating somewhat larger charge variations on the first and third layers, as compared to the second and fourth layers above the threshold doping value.

## S4.2 Tight Binding

The tight-binding model of the few-layer graphene is based on a single-particle $\vec{k}$-dependent tight-binding Hamiltonian with a unit cell comprised of two sites per layer, one for the A-sublattice and one for the B-sublattice of single-layer graphene, giving eight total sites for tetralayer graphene (see Fig. S11). The in-plane carbon-carbon and inter-layer distances are taken as $a = 1.42\,\text{Å}$ and $c = 3.4\,\text{Å}$, respectively.

We include on-site potential terms $\Delta$, $\Delta'$ on single (e.g. A2) and doubly (e.g. A3) eclipsed sites, respectively. Hopping parameters $\gamma_0$ through $\gamma_5$, defined in Fig. 1d of the main text, are taken into account. An in-plane next-nearest-neighbor hopping parameter, $t'$, has also been considered. However, it does not involve inter-sublattice mixing and was found to have a minor effect on the polarization. Thus, although it is present in the form of the Hamiltonian given below, practically it was set to zero.

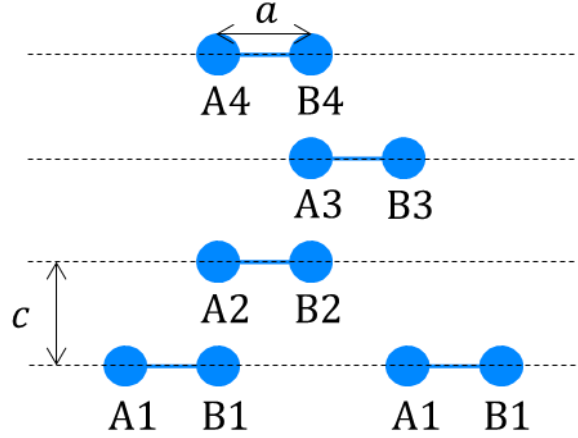

**Figure S11| Layout of the tight-binding model.**

The explicit $\vec{k}$-dependent Hamiltonian matrix for the four-layer polar configuration is given by:

$$\begin{pmatrix}
t'T(\vec{k}) & \gamma_0 S(\vec{k}) & \gamma_4 S(\vec{k}) & \gamma_3 S^*(\vec{k}) & 0 & \gamma_{2,R} & 0 & 0 \\
\gamma_0 S^*(\vec{k}) & \Delta + t'T(\vec{k}) & \gamma_1 & \gamma_4 S(\vec{k}) & 0 & 0 & 0 & 0 \\
\gamma_4 S^*(\vec{k}) & \gamma_1 & \Delta + t'T(\vec{k}) & \gamma_0 S(\vec{k}) & \gamma_4 S(\vec{k}) & \gamma_3 S^*(\vec{k}) & \gamma_{2,B} & 0 \\
\gamma_3 S(\vec{k}) & \gamma_4 S^*(\vec{k}) & \gamma_0 S^*(\vec{k}) & \Delta + t'T(\vec{k}) & \gamma_1 & \gamma_4 S(\vec{k}) & 0 & \gamma_5 \\
0 & 0 & \gamma_4 S^*(\vec{k}) & \gamma_1 & \Delta' + t'T(\vec{k}) & \gamma_0 S(\vec{k}) & \gamma_4 S^*(\vec{k}) & \gamma_1 \\
\gamma_{2,R} & 0 & \gamma_3 S(\vec{k}) & \gamma_4 S^*(\vec{k}) & \gamma_0 S^*(\vec{k}) & t'T(\vec{k}) & \gamma_3 S(\vec{k}) & \gamma_4 S^*(\vec{k}) \\
0 & 0 & \gamma_{2,B} & 0 & \gamma_4 S(\vec{k}) & \gamma_3 S^*(\vec{k}) & t'T(\vec{k}) & \gamma_0 S(\vec{k}) \\
0 & 0 & 0 & \gamma_5 & \gamma_1 & \gamma_4 S(\vec{k}) & \gamma_0 S^*(\vec{k}) & \Delta + t'T(\vec{k})
\end{pmatrix},$$

where $S(\vec{k})$ and $T(\vec{k})$ are given by:

$$S(\vec{k}) = 2e^{ia\frac{k_x}{2}} \cos\left(\frac{\sqrt{3}}{2} a k_y\right) + e^{-iak_x}$$

$$T(\vec{k}) = 2\cos(\sqrt{3} a k_y) + 4\cos\left(\frac{3}{2} a k_x\right) \cos\left(\frac{\sqrt{3}}{2} a k_y\right)$$

| Parameter | $\Delta$ | $\Delta'$ | $\gamma_0$ | $\gamma_1$ | $\gamma_{2,R}$ | $\gamma_{2,B}$ | $\gamma_3$ | $\gamma_4$ | $\gamma_5$ | $t'$ |
|---|---|---|---|---|---|---|---|---|---|---|
| Value (eV) | -0.008 | -0.016 | 3.16 | 0.39 | -0.02 | -0.02 | 0.315 | 0.044 | 0.038 | 0 |

**Table S2|** Numerical values of parameters used in the tight-binding model based on Ref. [11].

The Hamiltonian is then diagonalized to calculate the energy and the electron distribution across the unit cell sites of each state. In order to calculate the overall polarization for a given chemical potential and temperature, the different states are populated according to a Fermi-Dirac distribution. The electron distribution is then numerically integrated across the entire Brillouin Zone and all bands (accounting for spin degeneracy) to obtain the total electron occupation of each site. The polarization voltage in the out-of-plane direction is extracted by weighing each site occupancy by the corresponding vertical position from the center of the structure, assuming point-like orbitals.

**Self-consistent treatment of the Coulomb Interaction**

In order to account for the electrostatic interactions resulting from the non-uniform distribution of charge, a self-consistent approach is employed. Given a total charge distribution, it is possible to calculate the electrostatic potential differences between the different sites of the unit-cell. These potential differences may then be re-inserted into the single-particle Hamiltonian as on-site potentials. Since only potential differences have physical meaning, the inserted on-site potentials are always shifted such that they sum up to zero. The modified Hamiltonian is then solved to extract an updated charge distribution, with corresponding electrostatic potential differences. Using a nonlinear solver, a self-consistent set of potentials is found.

In general, the on-site potentials arising from the charge distribution can be expressed as $V_i = \sum_j G_{ij} n_j$, where $V_i$ is the on-site potential at site $i = A1, \cdots, B4$ (and similarly for $j$, see Fig. S11), $n_i$ is the corresponding excess charge, and $G_{ij}$ is an interaction matrix. Two different models were tested for the purpose of calculating the interaction matrix: A plate capacitor model and an Ewald summation model. Let us describe each in turn.

**Plate capacitor model:**

In the plate capacitor model, the overall charge of each layer, including the positive ions, is calculated based on the electron distribution. Each layer is treated as a uniformly charged capacitor plate of infinite dimensions, with a charge density equal to the total charge of a unit cell divided by its area. The potential differences between the different layers is then readily calculated. Explicitly, the interaction matrix for this model is given by:

$$G_{ij} = -|l(i) - l(j)| \frac{c}{2\varepsilon_0 A},$$

where $\varepsilon_0$ is the permittivity of free space, $A = \frac{3\sqrt{3}}{2} a^2$ is the unit-cell area, and $l(i) = 1, \cdots, 4$ is the layer index of lattice site $i$.

**Ewald Summation model:**

In this approach, the two-dimensional Ewald summation approach[12] is used to calculate the potentials in each lattice site, as generated by the net charge of all other sites in the infinite two-dimensional lattice, assuming point-like orbitals. The value of the cutoff parameter[12] was chosen to be $\xi = 0.4\sqrt{A}$, with $A$ being the unit cell area. We note, however, that different choices of this parameter had a negligible effect on the result.

Because Ewald summation is only valid when the net-charge of the unit cell is zero, any net charge of the distribution must be counteracted by uniformly adjusting the charge of the carbon ions such that the net charge is zero, as done also in the DFT calculations. We have verified that modelling the effect of the gate electrode by means of placing the extra charge on an external gate layer did not modify the results in any significant way.

Because the Ewald summation does not account for self-interaction energy within a lattice site, the chemical hardness of carbon[13] was modelled by adding a fixed value of $\eta = 10.0\ eV$ [13] on top of the electrostatic potential generated by all other lattice sites. Explicitly, the interaction matrix for this model is then given by:

$$G_{ij} = G_{ij}^0 + \eta \delta_{ij},$$

where $\delta_{ij}$ is the Kronecker delta and the numerical values of $G_{ij}^0$ are given (in units of [V/e]) by:

$$\begin{pmatrix}
-24.65 & -9.02 & -58.67 & -58.67 & -117.33 & -117.33 & -176.00 & -176.00 \\
-9.02 & -24.65 & -58.67 & -58.67 & -117.33 & -117.33 & -176.00 & -176.00 \\
-58.67 & -58.67 & -24.65 & -9.02 & -58.67 & -58.67 & -117.33 & -117.33 \\
-58.67 & -58.67 & -9.02 & -24.65 & -58.67 & -58.67 & -117.33 & -117.33 \\
-117.33 & -117.33 & -58.67 & -58.67 & -24.65 & -9.02 & -58.67 & -58.67 \\
-117.33 & -117.33 & -58.67 & -58.67 & -9.02 & -24.65 & -58.67 & -58.67 \\
-176.00 & -176.00 & -117.33 & -117.33 & -58.67 & -58.67 & -24.65 & -9.02 \\
-176.00 & -176.00 & -117.33 & -117.33 & -58.67 & -58.67 & -9.02 & -24.65
\end{pmatrix}.$$

For sites in different layers the values are very similar to the corresponding plate-capacitor ones. The main difference between the two models is in the intra-layer interactions, as discussed further below.

**Results and discussion**

In Figure S12 we compare the results of the different models for introducing a self-consistent electrostatic potential to the TB Hamiltonian: (i) none; (ii) plate capacitor model; and (iii) Ewald summation. When the self-consistent potential is neglected the polarization tends to be rather large. The self-consistent inclusion of the interaction strongly reduces the polarization value and may even flip its sign. The main difference between the plate capacitor and Ewald models is in the behavior of the polarization of the flat bands. Those bands, inherited from the Rhombohedral bottom 3 layers, reside mainly at the sites A1 and B3 [14]. At the K point and without the self-consistent potentials, only the $\gamma_{2,R}$ hopping amplitude is important, leading to bonding and antibonding eigenstates, the mean vertical position of which is close to layer 2. This behaviour holds also in the plate capacitor model. Only with Ewald summation, which allows for intra-layer potential difference, can the onsite potentials difference between A1 and B3 become sufficiently large with respect to $\gamma_{2,R}$. This then cause the eigenstates to reside mainly at one site or the other, as in the DFT calculation, which is important for obtaining both the right value of the polarization at charge neutrality and its doping dependence. Note that the trigonal warping caused by $\gamma_{3,4}$ leads to the gap between these two bands nearly closing away from the K point, a feature not seen in the DFT calculations, though without much effect on the polarization.

As for the low-energy Dirac bands, these are borne out of the Bernal top 3 layers. One of them mainly resides at site A4 and the other at an anti-bonding combination of B4 and B2 [14]. Their location and separation are thus determined by the combined effect of $\gamma_5$, $\Delta$, and the self-consistent electrostatic potential, and hence are quite sensitive to the parameter values. Only the Ewald model gives rise to a relatively symmetric bandstructure that is reminiscent of the DFT results.

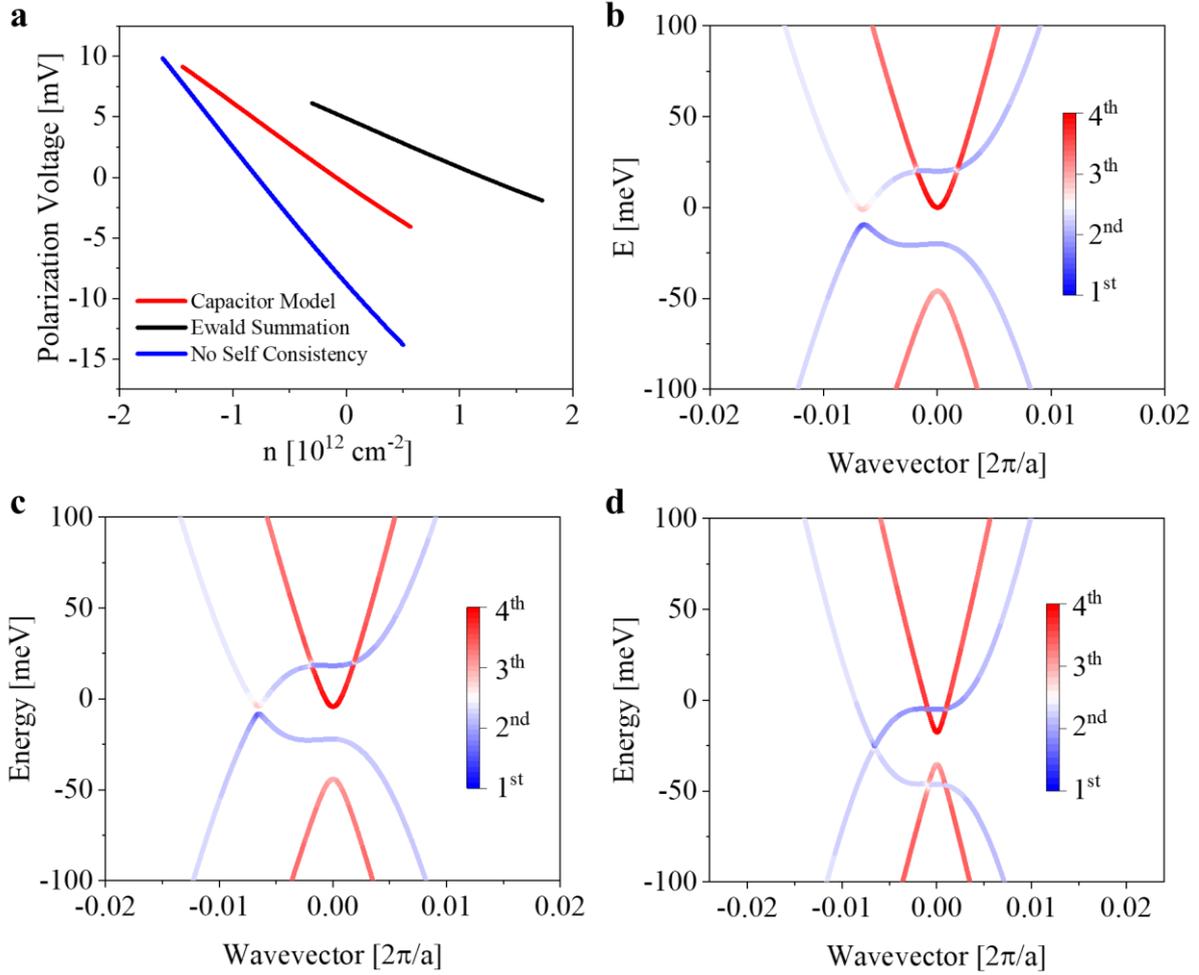

**Figure S12| Comparison of the different models of including electrostatic interactions in the TB calculation.** (a) Doping dependence of the polarization at room temperature. (b)-(d) near Fermi-level band structures, around the K point as in Fig. 3 of the main text. (b) No electrostatic interaction, (c) plate capacitor model, (d) Ewald summation model.